\documentclass[]{aa}

\usepackage{mwe}
\usepackage[utf8]{inputenc}
\usepackage[varg]{txfonts}
\usepackage[T1]{fontenc}
\usepackage{graphicx}
\usepackage{amsmath}
\usepackage{amssymb}
\usepackage{hyperref}
\usepackage{natbib}
\usepackage{hyperref}
\usepackage{enumerate}
\usepackage{adjustbox}
\usepackage{fancyref}

\usepackage[table]{xcolor}
\definecolor{gray75}{gray}{0.75}
\definecolor{dkgreen}{rgb}{0,0.6,0}
\definecolor{gray}{rgb}{0.5,0.5,0.5}
\definecolor{citationcolor}{rgb}{0,0.5,0}
\definecolor{linkcolor}{rgb}{0.5,0,0}

\newcommand*\mean[1]{\overline{#1}}

\hypersetup{colorlinks=true, breaklinks=true, linkcolor=linkcolor, menucolor=linkcolor, urlcolor=linkcolor,citecolor=citationcolor} 
\usepackage{longtable}
\usepackage[para]{threeparttable}
\usepackage[para]{threeparttablex}
\addtocounter{footnote}{1}
\numberwithin{equation}{section}
\usepackage{mathrsfs}
\usepackage{paralist}
\usepackage{url}
\usepackage{siunitx}
\usepackage{setspace}
\usepackage{multicol}
\usepackage{booktabs,makecell} 
\usepackage[]{float}
\usepackage{pdfpages}

\bibpunct{(}{)}{;}{a}{}{,}
\title{Ambilateral collimation study of the twin-jets in NGC\,1052}
\author{A.-K. Baczko\inst{\ref{affil:mpifr}}
\and E. Ros\inst{\ref{affil:mpifr}}
\and M. Kadler\inst{\ref{affil:wuerzburg}}
\and C.M. Fromm\inst{\ref{affil:Harvard2},\ref{affil:Frankfurt},\ref{affil:mpifr}}
\and B. Boccardi\inst{\ref{affil:mpifr}}
\and M. Perucho\inst{\ref{affil:dep_valencia},\ref{affil:dep_valencia2}}
\and T.P. Krichbaum\inst{\ref{affil:mpifr}}
\and P.R. Burd\inst{\ref{affil:wuerzburg}}
\and J.A. Zensus\inst{\ref{affil:mpifr}}
}
\institute{%
  Max-Planck-Institut f\"ur Radioastronomie, Auf dem H\"ugel 69, D-53121 Bonn, Germany
  \label{affil:mpifr}
  \and%
  Institut f\"ur Theoretische Physik und Astrophysik, Univ. W\"urzburg, Emil-Fischer-Str. 31, D-97074 W\"urzburg, Germany
  \label{affil:wuerzburg}
  \and%
  Black Hole Initiative at Harvard University, 20 Garden Street, Cambridge, MA 02138, USA
  \label{affil:Harvard2}
  \and%
  Institut für Theoretische Physik, Goethe-Universität Frankfurt, Max-von-Laue-Straße 1, D-60438 Frankfurt am Main, Germany
  \label{affil:Frankfurt}
  \and
  Dep. d'Astronomia i Astrof\'isica, Univ. de Val\`encia, C/ Dr. Moliner 50, E-46100 Burjassot, Val\`encia, Spain
  \label{affil:dep_valencia}
  \and%
  Observatori Astron\`omic, Univ. de Val\`encia, C/ Catedr\'atico Jos\'e Beltr\'an no. 2, E-46980 Paterna, Val\`encia, Spain
  \label{affil:dep_valencia2}
  }%
  
\begin{document}

 \abstract{
  \textit{Context:} With the increased sensitivity and resolution of radio interferometry the study of the collimation and acceleration region of extragalactic jets in Active Galactic Nuclei (AGN) has come into focus within the last years. Whereas a large fraction of AGN jets reveal a change from parabolic to conical collimation profile around the Bondi radius, there is a small number of sources which deviates from this standard picture, including the radio galaxy NGC\,1052. \newline
   \textit{Aims:} We study the jet width profile, which provides valuable information about the interplay between the central engine and accretion disk system and the formation, acceleration, and collimation of the jets. \newline
   \textit{Methods:} We observed the double-sided, low radio power active galaxy NGC\,1052 at six frequencies with the VLBA in 2017 and at 22\,GHz with RadioAstron in 2016. These data are combined with archival 15, 22, and 43\,GHz multi-epoch VLBA observations. From ridge-line fitting we obtained width measurements along the jet and counter-jet which were fitted with single and broken power-laws. \newline
   \textit{Results:} We find a clear break point in the jet collimation profile at $\sim 10^4\,R_\mathrm{S}$ (Schwarzschild radii). Downstream of the break the collimation is conical with a power-law index of $1.0$ -- $1.2$ (cylindrical 0; parabolic 0.5; conical 1) for both jets. On the other hand the upstream power-law index of $0.36$ for the approaching jet is neither cylindrical nor parabolic and for the receding jet with $0.16$ close-to cylindrical. For both jets we find a large opening angle of $\sim 30^\circ$ at a distance of $\sim 10^3\,R_\mathrm{S}$ and well collimated structures with an opening angle of $<10^\circ$ downstream of the break. \newline
   \textit{Conclusions:} There are significant differences in the upstream collimation profile between approaching (Eastern) and receding (Western) jet. Absorption or scattering in the surrounding torus as well as an accretion wind may mimic a cylindrical profile.  We need to increase the observing frequencies, which do not suffer from absorption to find the true jet collimation profile upstream of $10^4\,R_\mathrm{S}$.
 }
 \keywords{Galaxies:active --
           Galaxies:jets --
           High angular resolution
        }
               
 \maketitle

 \section{Introduction}
 \label{sec:intro}
	The most prominent features of radio-loud active galactic nuclei  \citep[AGN; see e.g.,][]{Zen97} are their relativistic jets \citep{Bla79}. To unveil the physical processes behind their formation and evolution, radio astronomical studies have focused on imaging the region around the central engine by means of high-frequency observations. To investigate the jet-forming region requires a high spatial resolution as well as a precise knowledge of the location of the central engine. The most famous example is \object{M\,87} which enabled through its close distance and high mass of the central supermassive black hole (SMBH) imaging of the jet base  very close to the central engine and its black hole shadow \citep{EHT19a,Kim18}. 
	
	Throughout the last years extensive studies of the acceleration and collimation region found a transition from a parabolic to conical collimation at distances of $\sim 10^4 - 10^6$ Schwarzschild radii ($R_\mathrm{S}$), where $R_\mathrm{S}= 2\,G\,M_{\rm BH}/c^2$. This is close to the gravitational sphere of influence of the black hole for several sources. This includes ten nearby sources of the \textsc{MOJAVE} sample and \object{M\,87} \citep{Kov20,Asa12}. There are several sources deviating from this standard picture. For example, in \object{NGC\,315} the transition is with $\sim 5\times 10^3$--$5\times10^4\,R_\mathrm{S}$ much closer to the central engine than the Bondi radius \citep{Boc21,Par21}. The high resolution provided by VLBI at millimetre wavelengths \citep[see ][]{Boc17}, in this case by the Global mm-VLBI array (GMVA), reveals very small transversal jet width \citep[e.g., M\,87,][]{Kim18}  as well as wide jet bases \citep[e.g., Cyg\,A, \object{3C\,84}][]{Gio18,Boc16b}. This suggests that both jet launching through the ergosphere of the black hole as well as by the accretion disk play an important role in the jet formation \citep{Bla82,Bla77}.
	
	In contrast to these sources, the twin-jet in \object{NGC\,1052} (PKS\,B0238$-$084, MCG\,$-$01$-$07$-$034, J0241$-$0815) does not show a parabolic collimation region. However, recent findings are controversial. On the one hand, by fitting the jet width at multiple frequencies at single epochs \cite{Nak20} found a cylindrical collimation with a power-law index of $k =0$ ($k =0.5$: parabolic; $k =1$: conical) at distances smaller than $10^4\,R_\mathrm{S}$, farther out the jets evolve conical. \object{3C\,84} is the only other source with a similar collimation profile \citep{Gio18}. On the other hand, based on a stacked 15\,GHz Very-Long-Baseline Array (VLBA) image \cite{Kov20} found a power-law index closer to a parabolic collimation profile for the upstream jet of $k=0.391\pm 0.048$. In this paper we will combine single-epoch and stacked, multi-frequency data of NGC\,1052 aiming at solving the puzzle of the upstream collimation profile.
	
    NGC\,1052 is classified as a low-ionization nuclear emission-line region object (LINER) based on its optical spectrum \citep{Fos78,Ho97}. It is located at a redshift of $z=0.005037$ \citep[assuming a systemic velocity of $v_\mathrm{sys} = 1507\,$\si{km/s};][]{Jen03}. Due to the close distance of the source, all our calculations are based on the redshift-independent distance of $D=19.23\pm0.14\,$Mpc \citep{Tul13}, leading to a linear scale of $0.093$ \si{pc/mas}. NGC\,1052 hosts a SMBH with a mass of $10^{8.2}\,M_\odot$ \citep{Woo02}. Recently, a significantly lower mass of $10^{5.51}\,M_\odot$ had been suggested assuming virialized BLR movements \citep{Kov20}. However, this method may be biased by optical obscuration of the central region. In addition, \cite{Kam20} infers a mass of $(2.13\pm0.09) \times 10^9 M_\odot$ inside the circumnuclear disk based on the measurement of the rotation curve of the circumnuclear disk. Considering this, and for consistency with our earlier work, we will assume a mass of $10^{8.2}\,M_\odot$ throughout the paper.
	
	NGC\,1052 is one of the few sources revealing a double sided. The radio structure inside the optical galaxy spans over about 3\,kpc and ends in diffuse lobes \citep{Kad04a}. At lower frequencies the intrinsic total radio power of this galaxy is a factor of 10--100 lower than the radio power of FR\,I radio galaxies, such as \object{M\,87} and \object{NGC\,315} (with radio structures larger than one Mpc). In comparison to powerful FR\,II radio galaxies, e.g., \object{Cygnus\,A}, its radio power is even about $10^4$ -– $10^5$ times lower. The symmetric radio jets in NGC 1052 are mildly relativistic ($\beta \lesssim 0.6$) in comparison to the highly relativistic jets in FR\,I and FR\,II radio galaxies. Based on several kinematic studies we adopt a viewing angle of $>80^\circ$ \citep{Ver03,Boe12,Bac19}. At centimetre wavelengths a torus with an optical depth of $\tau_{\rm 1\,GHz}\sim 300$ to $1000$ and a column density of $10^{22}\,\mathrm{cm}^{-2}$ to $10^{24}\mathrm{cm}^{-2}$ covers $\sim 0.1\,$pc towards the eastern, approaching jet and $\sim 0.7\,$pc towards the western, receding jet \citep{Kam01,Ver03,Kad04a,Kad04b,Saw08,Bre09}. This is consistent with a globally averaged column density of $(1-2) \times 10^{23}\mathrm{cm}^{-2}$ estimated from X-Ray observations \citep{Bal21}. This results in an emission gap between both jet cores at frequencies below 43\,GHz, see modeling by \cite{Fro18,Fro19}. At cm-wavelengths the mean speed of features in both jets have been derived as $\beta \leq 0.39$ \citep{Ver03,Boe12,Lis19}. Four years of 43\,GHz VLBA observations revealed higher apparent velocities of $\beta_\mathrm{ej}=0.53\pm0.04$ and $\beta_\mathrm{wj}=0.34\pm0.04$ for eastern and western jet, respectively, following an asymmetric evolution \citep{Bac19}. At 86\,GHz the jets have first been detected by the GMVA in 2004, revealing an nearly symmetric jet-counter-jet morphology with an unresolved central feature containing about 70\% of the total flux density \citep{Bac16}.  
	
	In this paper we focus on the collimation profile of NGC\,1052 derived from quasi-simultaneous VLBI observations at frequencies from 1.5\,GHz up to 43\,GHz. We combine the approach by \cite{Nak20} and \cite{Kov20} through inclusion  of quasi-simultaneous multi-frequency observations and stacked images at 15, 22, and 43\,GHz into our analysis. The multi-frequency observations and their data reduction is presented in \autoref{sec:observation}. In \autoref{sec:results} we give an overview of the methods used to derive the physical quantities, which will further be discussed in Section \autoref{sec:discussion}. Finally, \autoref{sec:summary} summarizes our results and gives an outlook.

 \section{Observation and Data reduction}
 \label{sec:observation}
 
 The frequency-dependent free-free absorption of the jet emission in the torus blocks our view onto the inner jet regions of NGC\,1052 at frequencies below 43\,GHz \citep{Kad04b}. On the other hand, higher frequency observations cannot image the extended, optically thin emission of the outer jet. Hence, in order to connect jet properties as collimation and opening angle from the parsec to the milli-parsec scales we observed NGC\,1052 at multiple frequencies from 1.5\,GHz up to 43\,GHz with the VLBA and a global array under the RadioAstron (RA) mission. In the following sections we will shortly summarize our individual observations, including data calibration and imaging.
 
\subsection{Multi-frequency VLBA observations}\label{sec:vlba}

We observed NGC\,1052 with the VLBA at the frequencies 1.5, 5.0, 8.4, 15, 22, and 43\,GHz on April 4, 2017. The data were recorded with the Roach Digital Backend (RDBE) at dual polarization with eight sub-bands per polarization band with a bandwidth of 32\,MHz and a data rate of 2\,Gbps. The correlation was performed at the National Radio Astronomy Observatory (NRAO) at the Array Operation Center of the VLBA in Socorro, NM, USA. In order to calibrate the \object{NGC\,1052} data the sources \object{0234+285}, \object{3C\,454.3}, and \object{4C\,39.25}  were observed.

\subsubsection{Calibration and imaging}
For calibration we applied dedicated procedures for the VLBA within the Astronomical Image Processing System 
($\mathcal{AIPS}$) \citep{Gre90}. Delay and phase offsets between the single intermediate frequencies (IF)s were removed through fringe detection on a strong calibration source. In order to perform a-priori amplitude correction the task \textsc{apcal} was applied using system temperatures provided by the array, this included corrections for the opacity. We performed global fringe fitting on all sources to solve for residual delays and rates. The calibrator source 3C\,454.3 was used to correct for the bandpass shape over frequency bands. After these calibration steps still persistent offsets between IFs were corrected by an additional round of fringe fitting. To enhance the fringe detection a preliminary image was produced in \textsc{difmap} \citep{She94}, which served as input model for the 
($\mathcal{AIPS}$) \textsc{fring} task while repeating the fringe fitting procedure for NGC\,1052. Following the calibration procedure data were averaged over frequency for each IF. To ensure the same calibration routine for all frequencies a Python script was applied, which automated the calibration process as far as possible. The script makes use of the \textsc{ParselTongue} python interface to 
$\mathcal{AIPS}$ \citep{Ket06}. 

\begin{figure*}[!htb]
   \centering
    \includegraphics{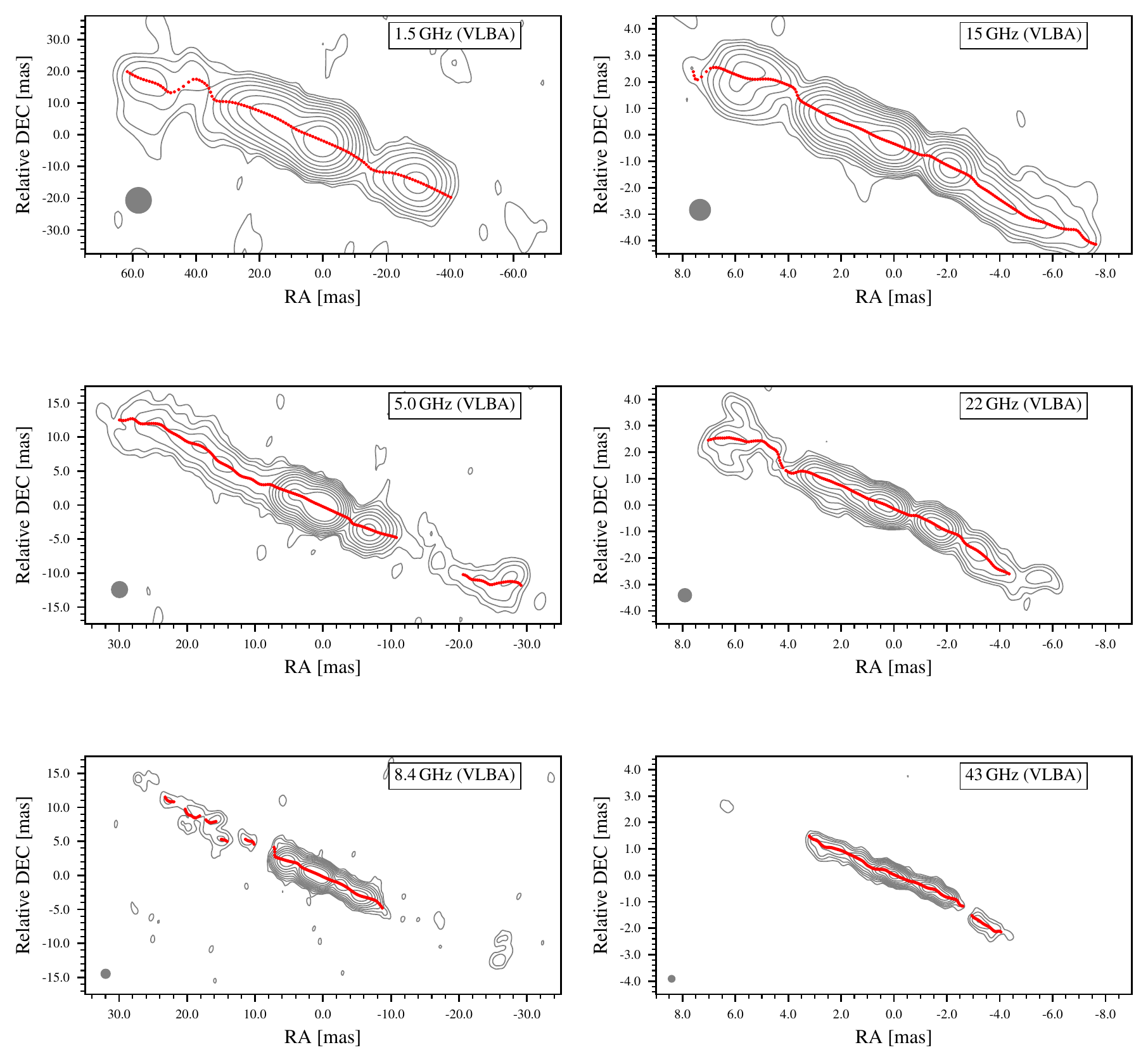}
    \caption{Final clean image of VLBA observations at all six frequencies restored with a circular beam (calculated as $b=\sqrt{b_\mathrm{maj}\times b_\mathrm{min}}$ with the major and minor beam according to table~\ref{tab:vlbaMFmapspars} column 7 and 8) with ridge-lines plotted on top (red dots).  The contours start at 4, 4, 4, 3, 2, and 3 times the noise level for $1.5$, $5.0$, $8.4$, $15$, $22$, $43$\,GHz (compare table~\ref{tab:vlbaMFmapspars} column 3), respectively and increase logarithmically by factors of 2.}
    \label{fig:vlbaMFrlOverplotcirc}
   \end{figure*}

Afterwards the data were read into \textsc{difmap} \citep{She94} for imaging and model fitting. Spurious data were flagged, followed by time averaging. Due the use of preliminary images during the  fringe detection in 
$\mathcal{AIPS}$, the phases were stable enough so we did not have to apply initial self-calibration with a point-source model in \textsc{difmap}. The final images (cf.\ Fig.~\ref{fig:vlbaMFrlOverplotcirc}) were produced using standard \textsc{clean} and self-calibration steps for phase and amplitude. The image parameters for the naturally weighted maps are listed in Tab.~\ref{tab:vlbaMFmapspars}. 

     \begin{table}[!hbt]
    \centering
    \small
    \setlength{\extrarowheight}{1pt}
    \setlength{\tabcolsep}{3pt}
    \caption[VLBA multi-frequency: \textsc{clean} image parameters.]{Image parameters for all analyzed VLBA observations from April 4, 2017 with natural weighting.}\label{tab:vlbaMFmapspars}
    \begin{adjustbox}{width=\linewidth}
    \begin{threeparttable}
      \begin{tabular}{lccccccccc}\toprule
	Array & $\nu$  & RMS\tnote{1a}  & RMS\tnote{1b}& $S_\mathrm{peak}$\tnote{2} &$S_\mathrm{tot}$\tnote{3} &$b_\mathrm{maj}$\tnote{4} &$b_\mathrm{min}$\tnote{5} & PA\tnote{6} & DR\tnote{7}  \\
	& $[$GHz$]$ &  $[\mathrm{\frac{mJy}{beam}}]$&  $[\mathrm{\frac{mJy}{beam}}]$& $[\mathrm{\frac{Jy}{beam}}]$&[Jy]&[mas]&[mas]&[$^\circ$] & \\\midrule
	VLBA & 1.5 	& 0.07 & 0.19 & 0.33 & 0.62 & 13.22 & 5.37 & $-5.9$ & 1737:1\\
	VLBA & 5.0	& 0.05 & 0.08 & 0.48 & 1.04 & 3.89 	& 1.65 & $-5.0$ & 6000:1\\
	VLBA & 8.4	& 0.07 & 0.10 & 0.36 & 1.12 & 2.30 	& 0.99 & $-5.0$ & 3600:1\\
	VLBA & 15.3	& 0.06 & 0.08 & 0.25 & 0.82 & 1.26 	& 0.55 & $-6.3$ & 3125:1\\
	VLBA & 22.2	& 0.08 & 0.09 & 0.18 & 0.80 & 0.86 	& 0.34 & $-5.0$ & 2000:1\\
	VLBA & 43.1	& 0.10 & 0.10 & 0.15 & 0.59 & 0.45 	& 0.19 & $-4.4$ & 1500:1\\\midrule
	RA-G\tnote{a} & 22.3& 0.04 & 0.04 & 0.11 & 0.42 & 0.44 & 0.39 & 44 & 1000:1 \\
	\bottomrule
      \end{tabular}
      \begin{tablenotes}\footnotesize
	\item[1] a) Root-mean-square (rms) noise level of image b) rms inside a structure-free window far away from the source structure
	\item[2] Peak flux density
	\item[3] Total recovered flux density
	\item[4,5,6] Major, minor axes and major axis position angle of the restoring beam
	\item[7] Dynamic range: ratio between the map peak and the rms inside a structure-free window far away from the source structure
	\item[a] RadioAstron ground-array image parameters
      \end{tablenotes}
    \end{threeparttable}
    \end{adjustbox}
  \end{table}   
   
   \subsection{RadioAstron observation}\label{sec:RA}
    
\begin{figure}[!hbt]
    \centering
    \includegraphics[width=0.9\linewidth]{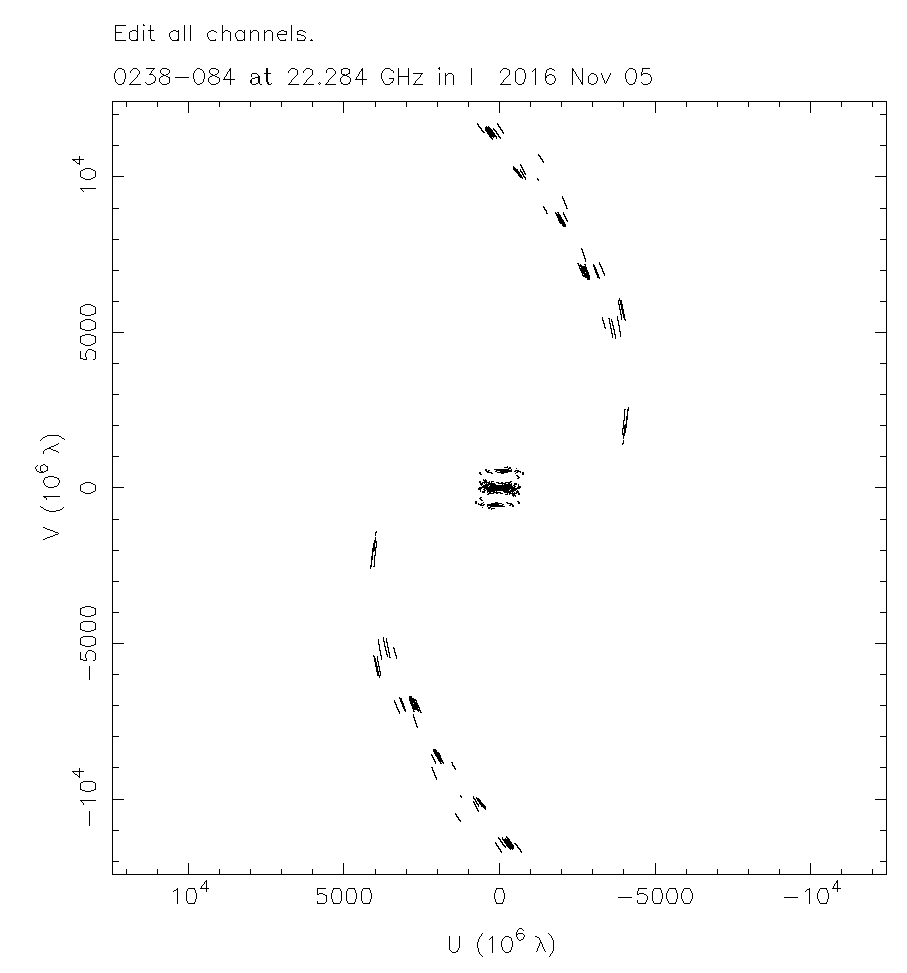}
    \caption{$(uv)$ coverage of the RadioAstron observation. We did not detect fringes on space-baselines.}
    \label{fig:RAuv}
   \end{figure}

    On November 5, 2017, \object{NGC\,1052} was observed with the 10-m Space Radio Telescope (SRT) \textit{Spekt-R} of the RadioAstron space-VLBI mission \citep{Kar12} and a large ground array consisting of antennas from the VLBA, Very Large Array (VLA), European VLBI Network (EVN), and the Long-Baseline Array (LBA) at 22\,GHz. The participation of the Sardinia radio telescope, Effelsberg, Green Bank Telescope, and phased Karl Jansky Very Large Array added high-sensitivity dishes to the array. The correlation was done at the DiFX software correlator in Bonn. The final set of telescopes was very heterogeneous, with different frequency setups. This resulted in different band widths and inconsistent number of band width for individual telescopes. To allow correlation of the experiment we used zoom-bands, which divides larger bands to match narrower bands and, hence, produce a consistent frequency band width for all telescopes \citep{Del11}. There was no significant fringe detection to the space antenna \textit{Spekt-R} at the Bonn correlator. Additional attempts by using the geodetic software package \textsc{PIMA} \citep{Pet11} did not improve the correlation. This can be explained either by 1) the large gap of $\sim 4$ earth diameter between ground-ground and ground-space baselines which reduces the SNR of possible fringes or 2) by the complex extended structure of NGC\,1052 at 22\,GHz at the projection angle sampled by the RadioAstron baselines, which might be resolved in space-ground baseline (cf.\ Fig~\ref{fig:RAuv}). Hence, the spatial scales sampled by the ground-ground baselines and the ground-space baselines are different. Therefore, solutions from the ground-ground fringe fitting cannot be applied to the long ground-space baselines and due to the very low SNR fringe fitting directly on the ground-space baselines was not possible. The final correlator output therefore includes the RadioAstron data without delay correction and twenty-nine antennas on ground.
   
   \subsubsection{Calibration and imaging}
   We followed a standard procedure to calibrate this data set using $\mathcal{AIPS}$, similar to our approach described in Sect.~\ref{sec:vlba} \citep[following a standard approach as described in e.g.,][]{Veg20}. Due to the large, heterogeneous array, the calibration had to be planned with care to not lose any visibility data. The final correlation output revealed a heterogeneous frequency coverage with a different number if IFs and  only half the full bandwidth for a subset of antennas. In order to correct for untypical slopes and offsets in the visibility amplitude we applied a bandpass calibration using the autocorrelated data. To improve the phase-calibration, a rough map of the data after initial calibration served as input model for an additional round of \textsc{fring} for the target source NGC\,1052. As a last step, the finally calibrated ground-array and the rough image was used as input to attempt fringe detection to the still uncorrected RadioAstron. However, also at this stage no fringes were detected. 
   
      \begin{figure}[!hbt] 
   \centering
 \includegraphics{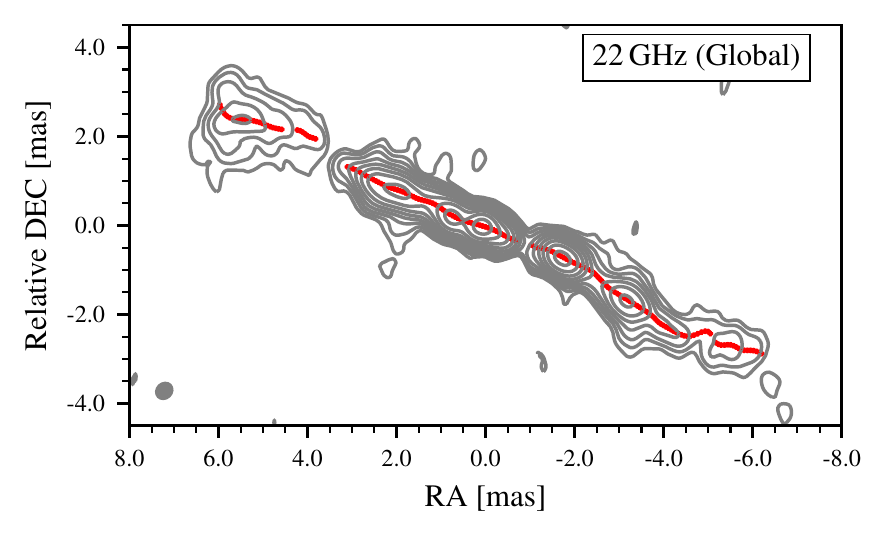}
 \caption[RadioAstron: \textsc{clean} images.]{Naturally weighted \textsc{clean} contour map of RadioAstron observation of NGC\,1052 at 22\,GHz in November\,2016, restored with a circular beam (calculated as $b=\sqrt{b_\mathrm{maj}\times b_\mathrm{min}}$ with the major and minor beam according to table~\ref{tab:vlbaMFmapspars} column 7 and 8). The ridge-line is plotted on top (red dots). Contour lines begin at four times the noise level (compare table~\ref{tab:vlbaMFmapspars} column 3) and increase logarithmically by factors of $2$.}\label{fig:RAimage}
\end{figure}
   Thanks to the large ground array this observation represents the highest resolution image at 22\,GHz of NGC\,1052 until now with a dynamic range of 1000:1. The inclusion of Southern stations enhanced the North-South resolution with a nearly circular beam of $0.443\times 0.392\,$(mas) at a position angle of \ang{44}. All parameters of the final \textsc{clean} map are listed in Table~\ref{tab:vlbaMFmapspars}. The final natural weighted clean map is shown in Fig.~\ref{fig:RAimage}.

    \subsection{Stacked VLBA observations}
    
\begin{figure}[!htb]
   \centering
    \includegraphics{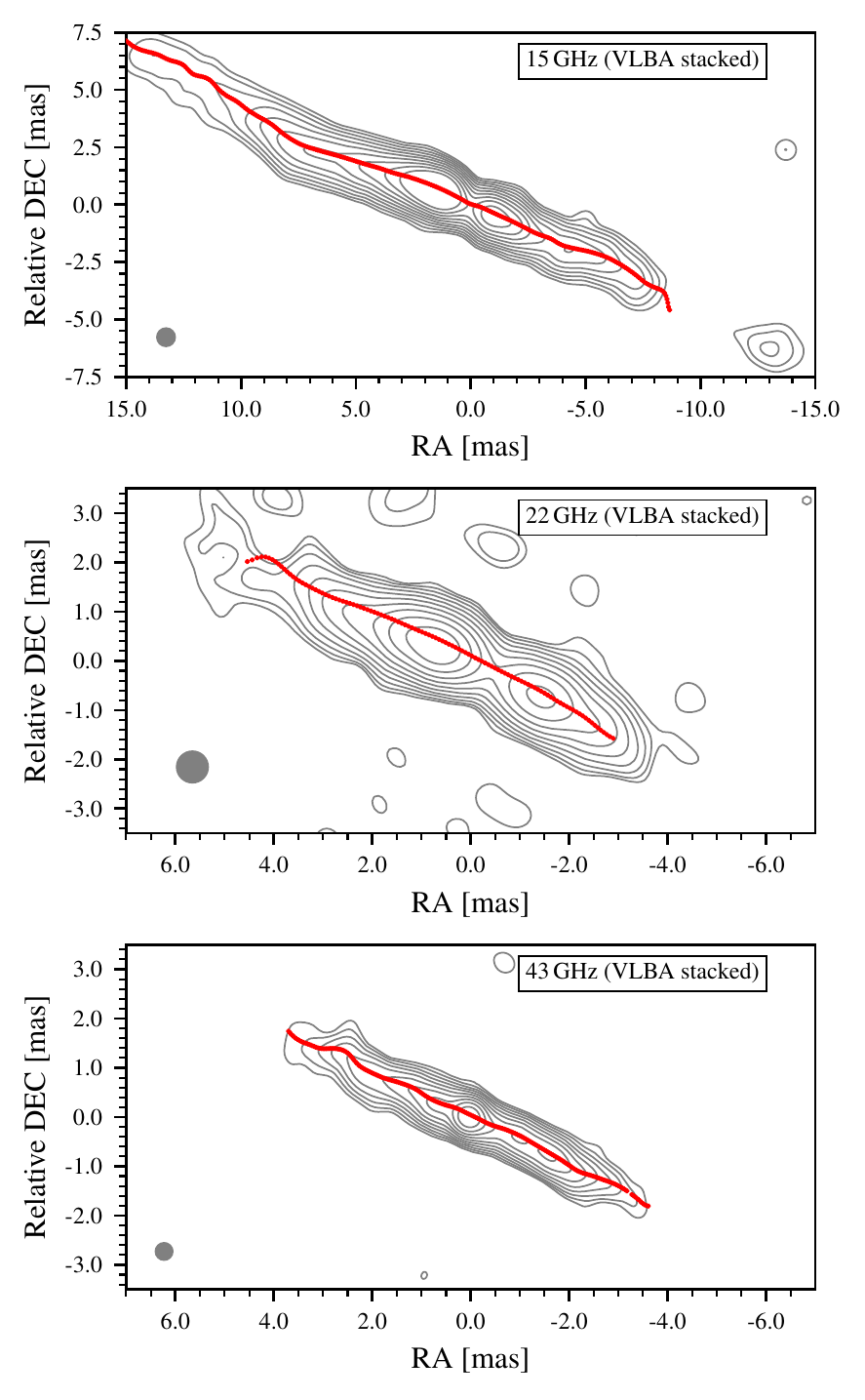}
    \caption{Stacked images based on four years of VLBA observations at 15\,GHz, 22\,GHz, and 43\,GHz \citep{Bac19,Lis19} restored with a circular beam. The ridge-lines is plotted on top (red dots).  The contours start at 0.5\,times the noise level for 15\,GHz, 22\,GHz, and 43\,GHz, which is derived as the mean of rms from the individual images, respectively and increase logarithmically by factors of $2$ (compare columns 2 and 5 of table~\ref{tab:stacked}).}
    \label{fig:vlbaSTrl}
   \end{figure}

    In addition to the new VLBA and RadioAstron observations we make use of four years of VLBA observations at 15\,GHz, 22\,GHz, and 43\,GHz from 2005 to 2009 of NGC\,1052, which have already been presented in \cite{Bac19} and \cite{Lis19}. To analyze the overall collimation profile along the twin-jet we produced stacked images at all three frequencies by restoring with a common circular beam of diameter $b = \sqrt{b_\mathrm{maj}\times b_\mathrm{min}}$. Based on the alignment presented in \cite{Bac19} the 22\,GHz images had been shifted with respect to the 43\,GHz images before deriving the mean image. In case of the archival 15\,GHz data the individual observations are not close enough to the 43\,GHz or 22\,GHz observations to align between frequencies. Assuming the center of the gap between both jet cores is very close to the central 43\,GHz peak, we aligned the 15\,GHz images on this position before estimating the mean image. The stacked images are shown in Fig.~\ref{fig:vlbaSTrl}. The total flux density of the RadioAstron ground image is significantly lower than the VLBA only image at the same frequency. This can be true variability as NGC\,1052 already showed strong variability of the total flux density within only a few months in previous observations. In addition, the RadioAstron observation suffered from several flux scaling problems. Significant changes of the visibility amplitude over the IFs adds a large uncertainty on the total flux density scaling of the image.
    
     \begin{table}[!hbt]
    \centering
    \caption[VLBA stacked.]{Image parameters for stacked VLBA maps.}\label{tab:stacked}
    \begin{threeparttable}
      \begin{tabular}{lcccc}\toprule
	 $\nu$  &  RMS\tnote{1} & $S_\mathrm{peak}$\tnote{2} &$S_\mathrm{tot}$\tnote{3} & beam\tnote{2}  \\
	$[$GHz$]$ &  $[\mathrm{\frac{mJy}{beam}}]$& $[\mathrm{\frac{Jy}{beam}}]$&[Jy]&[mas] \\\midrule
	15 & 0.23 & 0.36 & 1.07 & 0.87 \\
	22 & 0.73 & 0.36 & 1.01 & 0.67 \\
	43 & 0.63 & 0.34 & 0.79 & 0.39 \\
	\bottomrule
      \end{tabular}
     \begin{tablenotes}\footnotesize
        \item[1] The mean of the RMS from all images used in the stacking
        \item[2] Circular beam FWHM equivalent to the median of the common circular beams for each image
      \end{tablenotes}
    \end{threeparttable}
  \end{table}   
   
   \section{Results}
   \label{sec:results}
	\subsection{Image alignment of VLBA observations}\label{sec:align}
	
\begin{figure}[!hbt] 
     \centering
     \includegraphics[width=\linewidth]{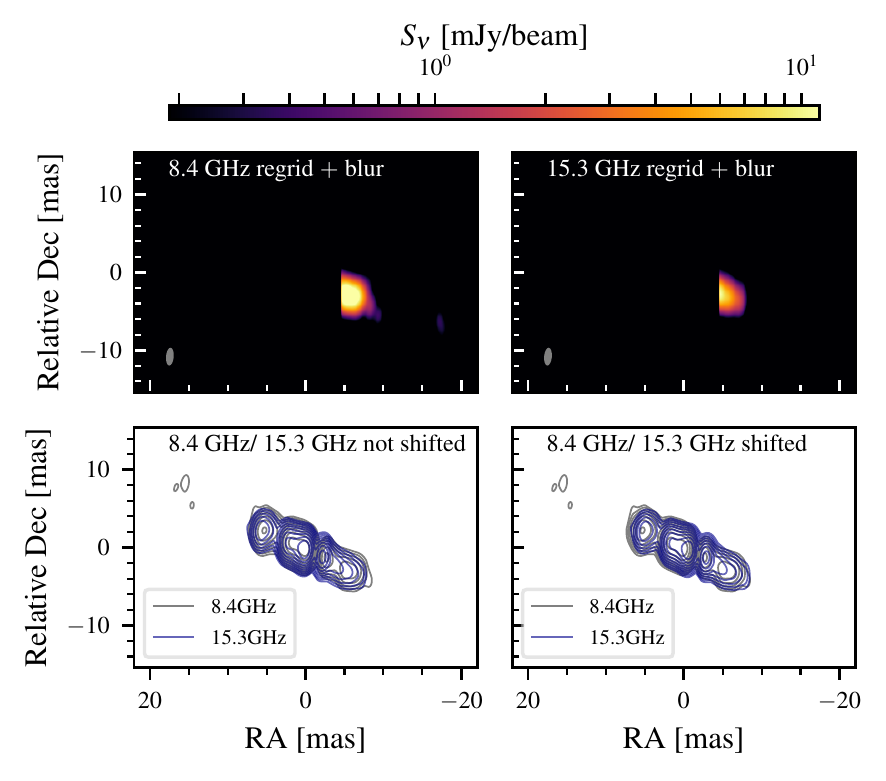}
     \caption[VLBA multi-frequency: alignment 8.4\,GHz-15\,GHz]{\textit{Top:} images of the \textsc{clean} model convolved with the natural restoring beam, with a mask applied for 8.4\,GHz (\textit{left}) and 15\,GHz (\textit{right}). Including the eastern jet emission resulted in an nonphysical small shift. Hence a mask was used to cut out the whole eastern jet and to only account for the emission of the western jet for the alignment. \textit{Bottom:} Contour maps without (\textit{left}) and with applied shift (\textit{right}).}
     \label{fig:vlbaMFalign1}
    \end{figure}

	We applied a 2D-cross-correlation in Fourier space, pairwise at adjacent frequencies in order to align all images \citep[making use of the \textsc{SCIKIT} and \textsc{EHTIM} packages][]{scikit,Cha16,Cha18}. The images were produced by the convolution of the \textsc{clean} component delta functions with the natural restoring beam of the lower frequency image on a grid with the pixel size and field of view of the higher frequency image. This approach ensures that the cross-correlation is most sensitive for the source structure and not the noise. Free-free absorption of a surrounding torus has a strong impact on the innermost region at frequencies below 43\,GHz. For the cross-correlation these areas have to be excluded to align only on optically thin structures. To account for this, we applied a mask to remove the obvious optically thick regions during the cross-correlation process. Figure~\ref{fig:vlbaMFalign1} visualizes the alignment. The upper panels show the re-gridded and blurred images at 8\,GHz and 15\,GHz with an elliptical mask applied. In the lower panels the contours of both maps are plotted without shift (left) and shifted according to the alignment (right). The derived pairwise shifts are listed in Table~\ref{tab:vlbaMFshifts}.
    \begin{table}[!ht]
     \caption[VLBA multi-frequency: shifts]{Shifts of frequency pairs resulting from 2D cross correlation on optically thin regions.}
     \label{tab:vlbaMFshifts}
     \centering
     \begin{tabular}{ccSS}
     \toprule
    {$\nu_\mathrm{low}\,[$GHz$]$} & {$\nu_\mathrm{high}\,[$GHz$]$} & {RA $[$mas$]$} & {DEC $[$mas$]$}\\
    \midrule
    $1.5$ & $5.0$ 	& -3.920 & 3.360 \\
    $5.0$ & $8.4$ 	& -1.080 & 0.720 \\
    $8.4$ & $15$ 	& -0.450 & 0.000\\
    $15$ & $22$ 	& -0.450 & 0.375 \\
    $22$ & $43$ 	& 0.152 & 0.076 \\
    \bottomrule   
     \end{tabular}
   \end{table}

\begin{figure}[!htb]
     \centering
     \includegraphics[width=\linewidth]{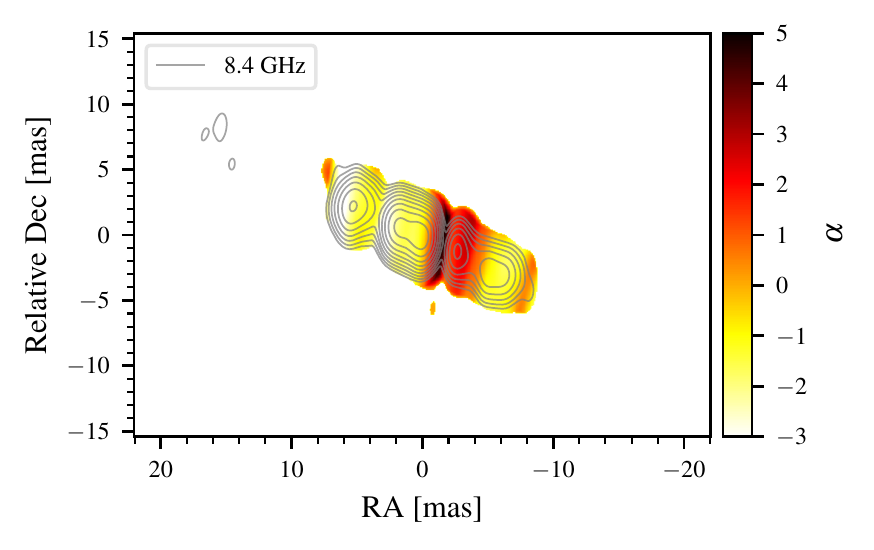}
     \caption[VLBA multi-frequency: spectral index maps.]{Spectral index map between 8.4\,GHz and 15\,GHz. The contours in the maps correspond to the lower frequency of the pair and start at three times the noise level.}
     \label{fig:vlbaMFspix}
    \end{figure}

	Previous observations of NGC\,1052 revealed highly inverted spectral indices at the location of the emission gap between both jet bases, exceeding a spectral index of $\alpha =2.5$ (where $S_\nu \propto \nu^\alpha$), which is explained by the obscuring torus resulting in free-free absorption. The same is observed for our new VLBA observations. Figure\,\ref{fig:vlbaMFspix} shows the spectral index map exemplary for the 8 and 15\,GHz images. Perpendicular to the jet axis the distribution of spectral indices is smooth and consistent with previous observations, which favours the correctness of our alignment. The same holds for the other frequency pairs.
  
     Based on these shifts all images have been aligned with respect to the 43\,GHz image. This approach assumes the brightest component at 43\,GHz to be the dynamic center of the source following \cite{Bac19}. These shifted images are used throughout the whole analysis when comparing properties of the jets at different frequencies.

  \subsection{The ridge--line in NGC\,1052}\label{sec:ResRidge}
   
   To derive the ridge-line, which connects the local flux density maxima along both jets, for each image we followed the subsequent steps for each frequency using the naturally weighted images which detect the outer jet structures with a high SNR: 1) The images had been restored with a circular beam of diameter $b= \sqrt{b_\mathrm{maj} \times b_\mathrm{min}}$ which maintains the beam area. All following procedures were run by a ParselTongue script (i.e., using 
   $\mathcal{AIPS}$ tasks): 2) Rotating the image by $24^\circ$ to align the main jet axis with the $x$-axis with the task \textsc{lgeom}; 3) Slicing these images perpendicular to the jet axis at each pixel with the task \textsc{slice}; 4) Fitting one Gaussian to the slices with \textsc{slfit}. From these fits we obtained the ridge-line as well as the jet width (equal to the deconvolved full-width half-maximum (FWHM) of the fitted slices) and peak flux density along the ridge-line. We include only fits with 1) a FWHM larger than the circular beam size and 2) a peak flux density higher than three times the noise level. The ridge-line is plotted on top of the clean images in Fig.~\ref{fig:vlbaMFrlOverplotcirc} and \ref{fig:RAimage}. The uncertainty on the deconvolved FWHM is estimated to be a combination of error propagation of the fitting error and the imaging error which we assumed as $1/10$ of the beam size \citep[cf.\ ][]{Par21}. This is reasonable due to the high dynamic range of our images and the limitation to width measurement with a signal-to-noise ratio (SNR) $\geq 3$. The uncertainty on the apparent opening angle was derived by error propagation.
   
\begin{figure*}[!htb]

	\centering
	\includegraphics[width=0.95\linewidth]{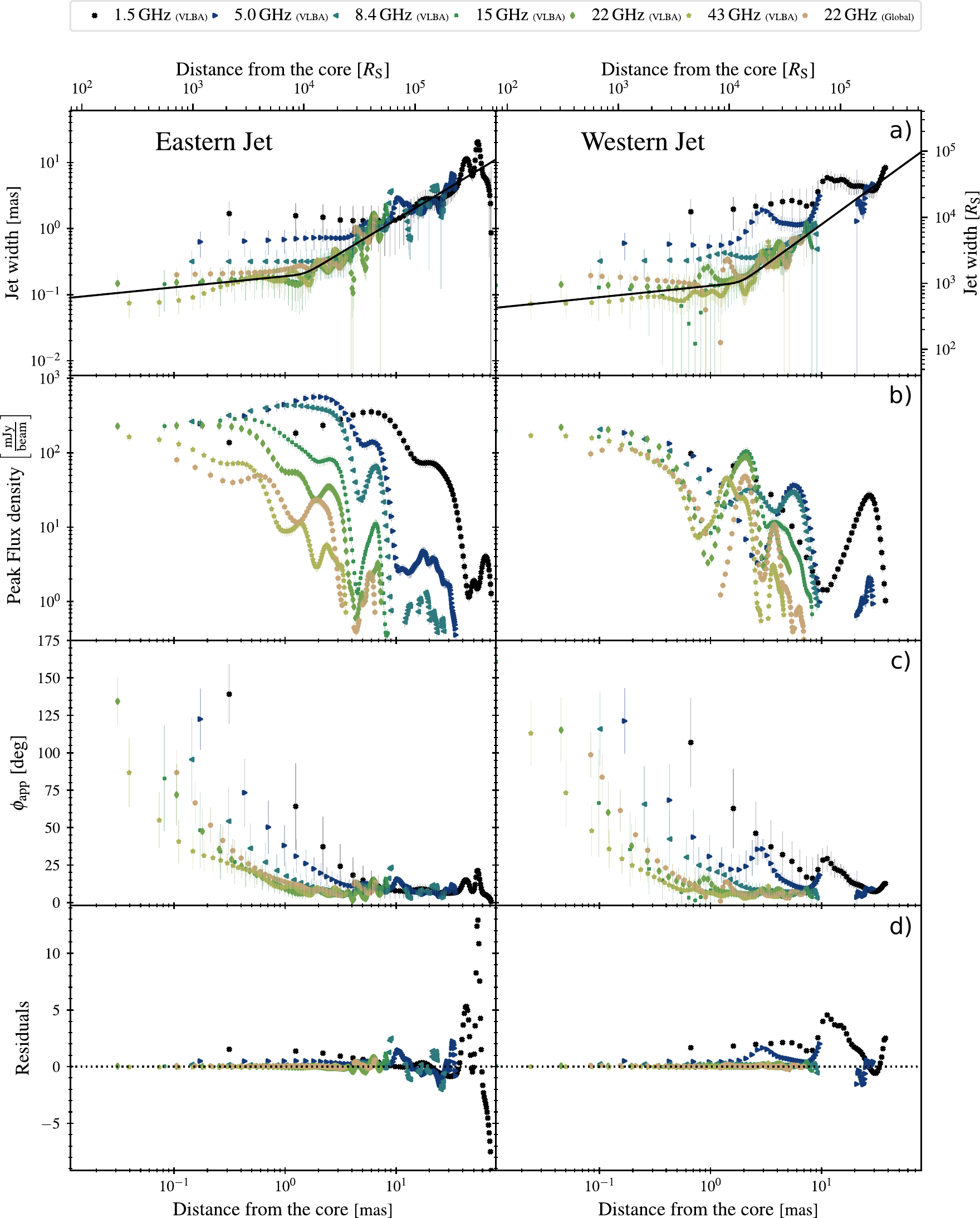}
	\caption{a) Jet width, b) peak flux density, c) apparent opening angle, and d) residuals of the data with the best fit along the jet ridge-line for Global (RadioAstron ground-support) and all six VLBA observations. Error bars are derived from a linear combination of the fitting and imaging error and are only plotted if exceeding the marker size. The values of $R_\mathrm{S}$ are computed assuming a black hole mass of $10^{8.2}~M_\odot$.}
	\label{fig:RAvlbawidth}
	\end{figure*}

   In Fig.~\ref{fig:RAvlbawidth} we compare jet width (top), peak flux density (middle top), apparent opening angle (middle bottom), and residuals (bottom) for all single-epoch observations for eastern jet (left) and western counter-jet (right). The ridge-lines were shifted according to Sec.~\ref{sec:align} with respect to the 43\,GHz image. We assumed the same shift for both 22\,GHz observations from the VLBA and RadioAstron. In case of the three stacked images the width and apparent opening angle are shown in Fig.~\ref{fig:vlbastackedwidth}.
   To examine the collimation profiles we fit power-laws to the jet width with distance. A first test of fitting a single power-law to both jets shows several deviations of the data from the model (compare Fig.~\ref{fig:vlba_singleRL}). At distances closer than $\sim 10^4\,R_\mathrm{S}$ the fitted power-law with index $k=1.05\pm0.03$ does not describe the measured width. Given this clear deviation from a single power-law we fit the data with a broken power-law model \citep[see e.g.,][]{Nak20}:
    \begin{equation}
     w(z) = W_0 2^{(k_\mathrm{u}-k_\mathrm{d})/s}\left(\frac{z}{z_\mathrm{b}}\right)^{k_\mathrm{u}}\left[1+\left(\frac{z}{z_\mathrm{b}}\right)^s\right]^{(k_\mathrm{d}-k_\mathrm{u})/s}.
    \end{equation}
    Here, $W_0$ is the width at the break point, $k_\mathrm{u/d}$ are the power-law indices for the upstream, and downstream parts ( $w \propto z^k$ ) 
    of the jet, that is, between dynamic center and break point and beyond the break point, respectively, and $z_\mathrm{b}$ is the break location. The parameter $s$ allows to define the degree of sharpness. The fitting revealed no significant dependence of the results for a sharpness between $10$ and $100$. In order to compare our results to \cite{Nak20} we fixed $s$ to a value of 10. 
    
    In addition, the Residuals in the lower panel in  Fig.~\ref{fig:vlba_singleRL} clearly indicate differences between the eastern and the western jet. Whereas the western jet at distances $> 10^4\,R_\mathrm{S}$ is fitted well, the eastern jet would require a steeper gradient of around $k\sim 1.2$. Hence, in the following we only discuss our results based on broken power-law fits to both jets individually. The fitting results are listed in Tab.~\ref{tab:widthfit} and Tab.~\label{tab:widthfitstacked}.

    \begin{table}[!ht]
    \centering
    \footnotesize
    \setlength{\extrarowheight}{1.5pt}
   \caption{Fitting results from a broken power-law fit to the jet width at all frequencies for the single epoch VLBA and RadioAstron observations.}
   \label{tab:widthfit}
   \centering
   \begin{adjustbox}{width=\linewidth}
   \begin{tabular}{@{}lcccc@{}}
    \toprule
    {Fit to} & {$W_0$} &{$k_\mathrm{u}$} & {$k_\mathrm{d}$} & {$z_\mathrm{b} \,[$mas$]$}\\
    \midrule
    Eastern Jet & 0.21 $\pm$ 0.02 & 0.17 $\pm$ 0.09 & 1.01 $\pm$ 0.01 & 1.51 $\pm$  0.19 \\
    Western Jet & 0.16 $\pm$ 0.01 & 0.17 $\pm$ 0.06 & 1.22 $\pm$ 0.02 & 1.88 $\pm$  0.15 \\
    Both Jets & 0.19 $\pm$ 0.02 & 0.18 $\pm$ 0.06 & 1.11 $\pm$ 0.01 & 1.73 $\pm$  0.14  \\
    \bottomrule
    \end{tabular}
    \end{adjustbox}
    \end{table}
    
   \section{Discussion}
   \label{sec:discussion}
		 
   \subsection{Collimation profiles}\label{sec:jetwidth}
   
\begin{figure}[!htb]
	\centering
	\includegraphics{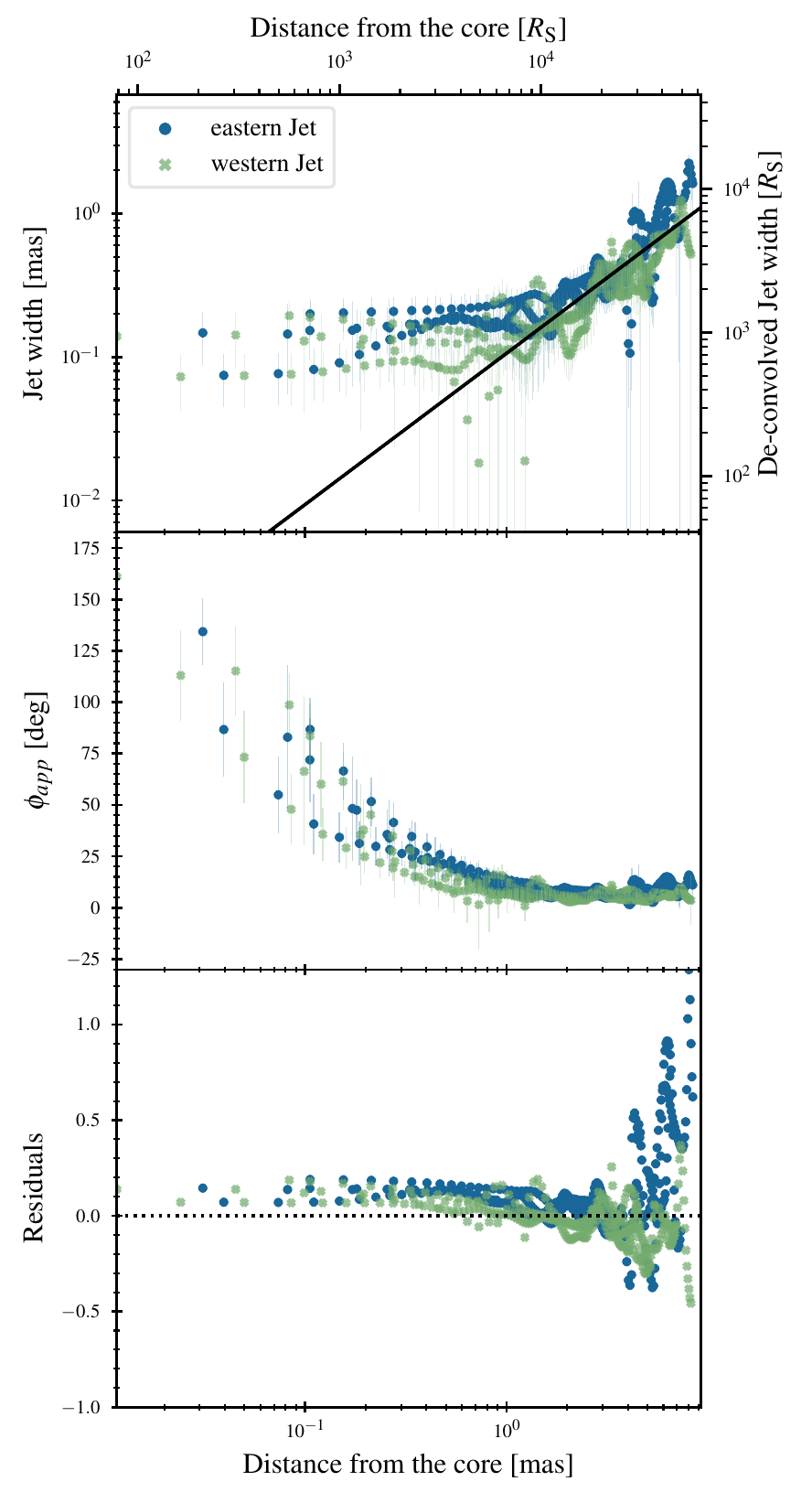}
	\caption{(\textit{Top}) Jet width, (\textit{Middle}) apparent opening angle, and (\textit{Bottom}) residuals along jet ridge-line for VLBA 15, 22, and 43\,GHz and Global 22\,GHz observations. The black line in the upper panel shows the best fit for a single power law. Error bars are derived from the fitting error and imaging error of $1/10$ of the beam size and are only plotted if exceeding the marker size. The values of $R_\mathrm{S}$ are computed assuming a black hole mass of $10^{8.2}~M_\odot$.}\label{fig:RAvlba_u-q_Powerlaw}
	\label{fig:vlba_singleRL}
	\end{figure}

   In order to study the collimation profile of the jets in NGC\,1052 we fitted a broken power-law to the width along both jets individually, as described in Sect.~\ref{sec:ResRidge} (cf.\ Fig.~\ref{fig:RAvlbawidth}). Assessing the goodness of the fit is difficult due to a number of reasons. The fitted model is non-linear. Moreover, as we took measurements of the width of the jet in each pixel, which is smaller than the beam size, neighboring measurements are likely correlated. Hence, the reduced $\chi^2$ does not provide a reliable method for estimating the goodness of the fit. Therefore, we assess the residuals, plotted on the bottom panel of Fig.~\ref{fig:RAvlbawidth}, to evaluate the goodness of the fit. Based on this approach we favour the broken power-law. The single power-law shows a clear discrepancy of the data points from the fit at distances $<2\,$mas to the center (cf.\ Fig.~\ref{fig:RAvlba_u-q_Powerlaw} as an example for a single power-law fit to both jets). 
   
   We also cross checked the smoothly broken power law fit with the following procedure. For both jets we generated $10^4$ random break points along the entire jet length drawn from all frequencies. For each break point we applied a single power law fit in the inner and outer jet region. After all runs we applied quality filters for the fits $0.8\leq X^2_\mathrm{red} \leq 3$ and prohibited cases where the inner or outer region would be fit with more than 80\% of the data along the jet length. From the remaining we calculated the median and the mean for both jets (see Table~\ref{tab:quality_check}), which is consistent with the findings for the smoothly broken power law.
   
    \begin{table}[!ht]
    \centering
    \setlength{\extrarowheight}{1.5pt}
    \caption{Collimation profile results from cross check.}
    \label{tab:quality_check}
    \centering
    \begin{adjustbox}{width=\linewidth}
    \begin{tabular}{lcccccc}
    \toprule
    {Fit to} &{$\mean{k_\mathrm{u}}$} &{$k_\mathrm{u}^\mathrm{med}$} & {$\mean{k_\mathrm{u}}$} &{$k_\mathrm{u}^\mathrm{med}$} & {$\mean{z_\mathrm{b}} [$mas$]$}& {$z_\mathrm{b}^\mathrm{med} [$mas$]$}\\
    \midrule
    Eastern Jet & 0.96 & 0.96 & 0.29 & 0.18 & 1.77 & 1.12  \\
    Western Jet & 0.92 & 0.93 & 0.26 & 0.16 & 2.0 & 1.74 \\
    \bottomrule
    \end{tabular}
    \end{adjustbox}
    \end{table}
   
   There are significant differences between eastern and western jet, especially at 1.5\,GHz and 5\,GHz. Hence, we only discuss our results from fitting the width of each jet individually. 
   
   Downstream of the break point at around $10^4\,R_\mathrm{S}$ the jets evolve quasi conical. Upstream they are close to cylindrical with a power-law index of $k_\mathrm{u}\sim 0.17$. This kind of break in the collimation profile has been found for a number of (mostly one-sided) AGN jets. This includes low-luminosity AGN as M\,87 and 3C84, as well as high-luminosity radio galaxies as, e.g., Cygnus~A. Assuming that the jet phenomenon is indeed universal, we expect to observe a similar behaviour for AGN jets with different radio powers. This makes a comparison of NGC\,1052 to other low-power and high-power sources extremely valuable for our understanding of the underlying physical processes behind the formation and collimation of jets at all scales.
   
   For ten sources out of the MOJAVE 15\,GHz sample \cite{Kov20} found a break point in a typical range of $r_\mathrm{break} \in (10^5,10^6)r_\mathrm{S}$, at which the shape changes from parabolic to conical. The same region was found in a few single-sources studies, including \object{NGC\,6251} \citep{Tse16} and \object{M\,87} \citep{Kim18}. Within this comparison, NGC\,1052 stands out by not showing a parabolic expansion. One of the very few other sources with a similar cylindrical geometry is \object{3C\,84} with a collimation profile of $r\propto z^{0.17\pm0.08}$ at distances of $180\,R_\mathrm{S}<z<8000\,R_\mathrm{S}$, measured with RadioAstron \citep{Gio18}. \cite{Nag14} supports a cylindrical profile down to at least $5\times 10^4\,R_\mathrm{S}$ for this source. Similar wide, but parabolic evolving jet bases are found in \object{Cyg\,A} with a minimum jet width of $227\pm98\,R_\mathrm{S}$ \citep{Boc16b}. In section~\ref{sec:colac} we will further discuss the impacts of these findings on the collimation and acceleration zone.

\begin{figure*}[!htb]
	\centering
	\includegraphics{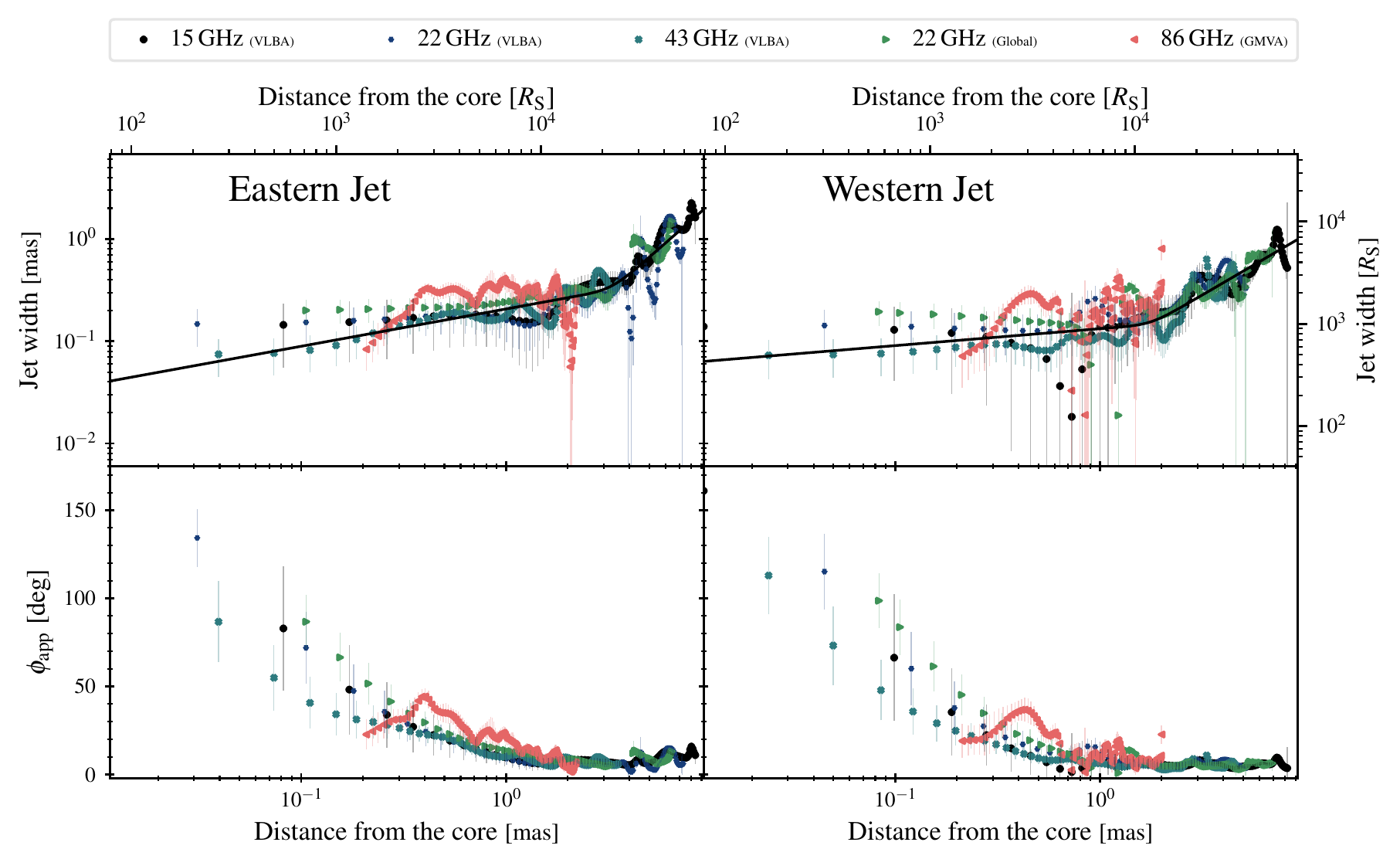}
	\caption{Jet width (top) and apparent opening angle (bottom) along the jet for Global 22\,GHz and VLBA 15, 22, and 43\,GHz observations. The same parameters for the GMVA 86\,GHz observation from 2004 are plotted on top, without being taken into account for the fit. Error bars are derived from a linear combination of the fitting and imaging error and are only plotted if exceeding the marker size. The values of $R_\mathrm{S}$ are computed assuming a black hole mass of $10^{8.2}~M_\odot$.}
	\label{fig:RAvlba_u-w_width}
	\end{figure*}

   The low frequency data in the western jet are not consistent with a power-law as the width is not simply decreasing upstream of the jet, but follows a step pattern. This cannot be observed in the eastern jet. We can exclude that this is due to an artefact of our analysis, as the slice profile clearly shows one peak and is fitted accurately with a single Gaussian. Similar larger width in the western jet were also found by \cite{Nak20}. Hence, we assume that this is a common feature observable in NGC\,1052 and not due to the morphology of the source at the time of the observation. Assuming this effect is not intrinsic to the jet, interaction with the surrounding medium is a likely explanation. Several authors found a dense plasma torus which covers larger parts of the western jet with respect to the eastern jet \citep[see e.g.,][]{Kam01,Kad04a,Saw08}. Coinciding with the torus $\mathrm{H_2O}$ maser emission was found \citep[see e.g.,][]{Cla98}. In addition, recent ALMA observations at sub-mm wavelengths found a massive molecular torus with $M_\mathrm{torus}=(1.3\pm0.3)\times 10^7\,M_\odot$ inside a gas-poor circumnuclear disk \citep{Kam20}. With a radius of $2.4\pm 1.3\,$pc and a thickness ratio of $0.7\pm0.3$ it covers the part of the western jet with deviating width measurements. Previous observations already connect the excess of absorption in between the jet cores with free-free absorption in a surrounding plasma torus \citep{Kam01}. With increasing observing frequency its impact gets smaller and at around 43\,GHz is nearly negligible \citep{Bac19}. This absorption effect is larger for the western jet. With respect to the molecular torus we not only may expect absorption but in addition scattering of the jet emission. As both effects are frequency dependent, this explains the large deviation of the widths at $\nu<15\,$GHz from a power-law. 
   To exclude this effect we will further focus on the collimation profile derived from our images with observing frequencies $\geq 15\,$GHz. The corresponding best-fit result is show in Fig.~\ref{fig:RAvlba_u-w_width} and Table~\ref{tab:widthfit15Ghz}. 
    \begin{table}[!ht]
    \centering
    \setlength{\extrarowheight}{1.5pt}
    \caption{Fitting results from a broken power-law fit to the jet width at frequencies $\geq 15\,$GHz.}
    \label{tab:widthfit15Ghz}
    \centering
    \begin{adjustbox}{width=\linewidth}
    \begin{tabular}{lcccc}
    \toprule
    {Fit to} & {$W_0$} &{$k_\mathrm{u}$} & {$k_\mathrm{d}$} & {$z_\mathrm{b} [$mas$]$}\\
    \midrule
    Eastern Jet & 0.35 $\pm$ 0.02 & 0.36 $\pm$ 0.05 & 1.76 $\pm$ 0.12 & 3.30 $\pm$ 0.24\\
    Western Jet & 0.16 $\pm$ 0.01 & 0.16 $\pm$ 0.06 & 1.17 $\pm$ 0.04 & 1.78 $\pm$ 0.17\\
    Both Jets & 0.32 $\pm$ 0.02 & 0.42 $\pm$ 0.04 & 1.70 $\pm$ 0.09 & 3.45 $\pm$ 0.19\\
    \bottomrule
    \end{tabular}
    \end{adjustbox}
    \end{table}
    
	The residuals still show deviations at larger distances. However, their amplitude is smaller and they spread symmetrically around $0$. Due to the absorbing effect of the surrounding torus our data are sparse upstream of the break point and contain larger uncertainties in the western jet. There are now significant differences of the power laws between eastern and western jet. In the approaching, eastern jet the break point of $3.3\,$mas and the upstream power-law index of $k_\mathrm{u}=0.36\pm0.05$ is consistent with the findings of \cite{Kov20}, whose measurements are based on 15\,GHz VLBA data from the eastern jet. As the break point is found to be relatively far outside, the downstream power-law index is fitted with higher accuracy using all frequencies. Hence, from Table~\ref{tab:widthfit} and Fig.\ref{fig:RAvlbawidth} this results in a downstream power-law index for the eastern jet of $k_\mathrm{d}=1.01\pm0.01$. The western jet changes from a more collimated jet to a conical expansion at a closer distance of $1.8\,$mas.
	
	Further, our results are comparable to the findings by \cite{Nak20} who derived the collimation parameters together for both jets based on archival observations. However, their work suggests a flatter profile close to $k_\mathrm{u}=0$. In addition, they did not find differences between both jets, but suggest a symmetrical evolution. This discrepancy may be explained by the higher sensitivity of our data set in comparison to the older, archival data which were obtained in 2000 and 2001. Multi-epoch VLBA data at 43\,GHz revealed changes in the morphology of the source within four years, being symmetrical in the beginning, and developing an asymmetric structure within one year. This resulted in the western counter-jet being brighter than the eastern jet \citep{Bac19}. Therefore, it cannot be excluded that the morphology of the source in our observations is different, also in terms of jet collimation, in comparison to the data from 2001. 
	
\begin{figure*}[!hbt]
	\centering
	\includegraphics[width=0.98\linewidth]{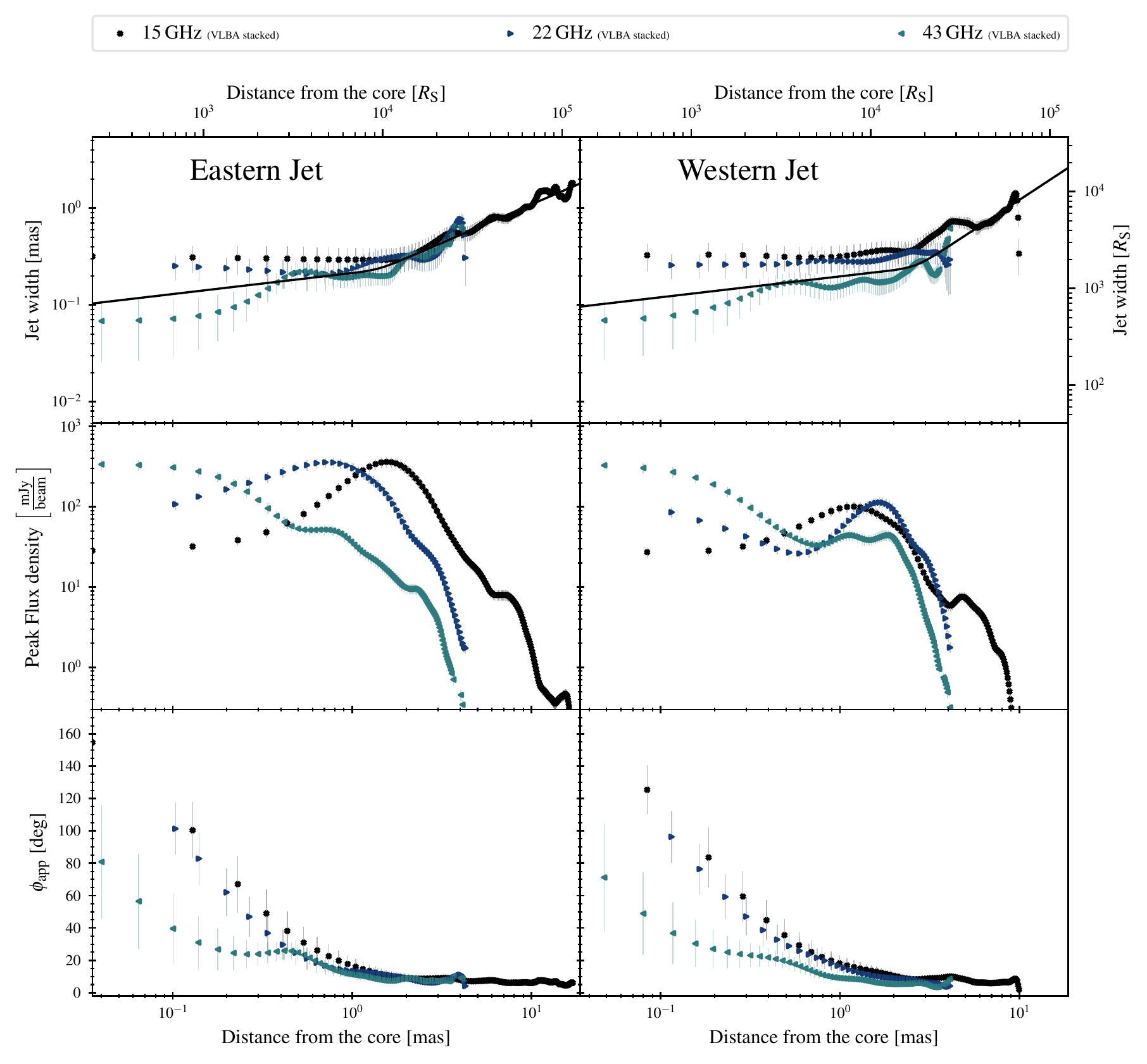}
	\caption{\textit{Top:} Jet width, \textit{Middle:} peak flux density, and \textit{Bottom:} apparent opening angle of the data with the best fit along jet ridge-line for stacked 15\,GHz, 22\,GHz, and 43\,GHz VLBA images. Error bars are derived from a linear combination of the fitting and imaging error and are only plotted if exceeding the marker size. The values of $R_\mathrm{S}$ are computed assuming a black hole mass of $10^{8.2}~M_\odot$.}
	\label{fig:vlbastackedwidth}
	\end{figure*}

	To study whether these differences between both jets are a common feature or specific at this time of observation, in the following, we will discuss the outcomes from fitting the stacked VLBA images at 15, 22, and 43\,GHz. In comparison to the higher-frequency single-epoch images (Fig.~\ref{fig:RAvlba_u-w_width}), the fitting to the stacked images (Fig.~\ref{fig:vlbastackedwidth}) reveals consistent results for both jets, with a close-to cylindrical upstream power-law index of $k_\mathrm{u}= 0.21$ and a conical downstream power-law index of $k_\mathrm{d}=0.80$ and $k_\mathrm{d}=1.22$ for eastern and western jet, respectively. The widths of the western jet at 22\,GHz and 15\,GHz are slightly wider in comparison to the eastern jet. This points towards scattering at 15\,GHz and 22\,GHz in the western jet. In the eastern jet, the 15\,GHz and 22\,GHz width upstream of $\sim 10^4\,R_\mathrm{S}$ do not decrease with closer distance to the core but follow a clear cylindrical profile. At distances closer than $0.3\,$mas the widths measured at 43\,GHz deviate from this and clearly decrease. Fitting only the width of the 43\,GHz stacked image at distances $z<2\,$mas results in power-law indices of $k_\mathrm{u}=0.30\pm 0.03$ and $k_\mathrm{u}=0.22\pm0.02$ for eastern and western jet, similar to the fitting results obtained for the single-epoch observations. This difference between 43\,GHz and lower frequencies hint towards different layers of the jet being responsible for the emission at the individual frequencies. In Sec.~\ref{sec:colac} we will shortly discuss the possibility of differential expansion.
	
	\begin{table}[!ht]
    \centering
    \setlength{\extrarowheight}{1.5pt}
    \caption{Fitting results from a broken power-law fit to stacked 15, 22, and 43\,GHz VLBA images.}
    \label{tab:widthfitstacked}
    \centering
    \begin{adjustbox}{width=\linewidth}
    \begin{tabular}{@{}lcccc@{}}
    \toprule
    {Fit to} & {$W_0$} &{$k_\mathrm{u}$} & {$k_\mathrm{d}$} & {$z_\mathrm{b}\, [$mas$]$}\\
    \midrule
    Eastern Jet & 0.24 $\pm$ 0.02 & 0.22 $\pm$ 0.06 & 0.80 $\pm$ 0.01 & 1.40 $\pm$ 0.18\\
    Western Jet & 0.26 $\pm$ 0.02 & 0.21 $\pm$ 0.05 & 1.22 $\pm$ 0.05 & 2.66 $\pm$ 0.22\\
    Both Jets & 0.24 $\pm$ 0.02 & 0.20 $\pm$ 0.05 & 0.90 $\pm$ 0.01 & 1.79 $\pm$ 0.15\\
    \bottomrule
    \end{tabular}
    \end{adjustbox}
    \end{table}

	\subsection{The advantage of stacked images}
	
	Previous findings on the collimation profile of the jets in NGC\,1052 are controversial. While multi-frequency, but single-epoch observations found a cylindrical collimation profile with $k_\mathrm{u}=0$ upstream of $10^4\,R_\mathrm{S}$ \citep{Nak20}, the power-law index based on a $15$\,GHz stacked image is with $k_\mathrm{u}=0.391\pm 0.048$ closer to a parabolic shape \citep{Kov20}. On the one hand, \cite{Kov20} only analyzed the profile for the approaching jet and hence, no study about the symmetry of the source was made. On the other hand, \cite{Nak20} found both jets to be symmetric within errors, however the sampling of width measurements upstream of the break point is sparse and the broken power-law fit is largely guided by large scale structures. 
	
	In our analysis we included the same VLBA 15\,GHz images as \cite{Kov20} applying a slightly different approach for alignment. Whereas the stacked images used in \cite{Kov20} are aligned on the VLBI core, we shifted all images on the center of the emission gap between both cores, where the $43$\,GHz core is assumed to reside \citep{Bac19}. This allows us to make a combined analysis of the 15, 22, and $43$\,GHz stacked images. Furthermore, the herein presented multi-frequency VLBA observations have an improved sensitivity in comparison to the observations used by \cite{Nak20}, which were obtained more than ten years earlier. This results in detection of fainter jet emission in the outer parts of the jets and with that in more pronounced differences between both jets. 
	
	With upstream power-law indices of $0.16 \leq k_\mathrm{u} \leq 0.35$ we cannot confirm a truly cylindrical profile as proposed by \cite{Nak20}, but taking into account only observations at frequencies $\nu\geq 15\,$GHz the power-law profile of the approaching, eastern jet is consistent with the findings by \cite{Kov20}. However, the stacked $43$\,GHz width profile indicates a more complicated structure, with a conical profile at $z\geq 10^4\,R_\mathrm{S}$, a cylindrical profile at $3\times 10^3\,R_\mathrm{S}\leq z\leq 10^4\,R_\mathrm{S}$, and a quasi-parabolic profile at $z\leq 3\times 10^3\,R_\mathrm{S}$. 
	
	A comprehensive study of multi-epoch 43\,GHz observations of NGC\,1052 revealed a dramatic change of the source morphology from being symmetric to asymmetric within time scales of about one year \citep{Bac19}. This suggests intrinsic asymmetric ejection of jet plasma into both the approaching and receding jets. As a consequence the collimation profile might change depending on the observation date. This could explain the discrepancy of our multi-epoch observations to the findings by \cite{Nak20}. These temporal asymmetries of single-epoch observations make studies of jet collimation difficult. In contrast to that our results from the multi-frequency stacked images suggest, that the averaging allows us to measure the width of the whole jet channel, and hence, the true collimation profile, which can be connected to physical models on jet formation. Hence, in the case of NGC\,1052 these kind of studies have to be made with stacked images, especially at higher frequencies, to ensure that the whole jet channel is included in the analysis. Only then, reliable statements about the large-scale jet physics can be made.
	
	\subsection{Comparison to 86\,GHz GMVA observations}
	
\begin{figure}[htb]
       \centering
       \includegraphics[width=0.98\linewidth]{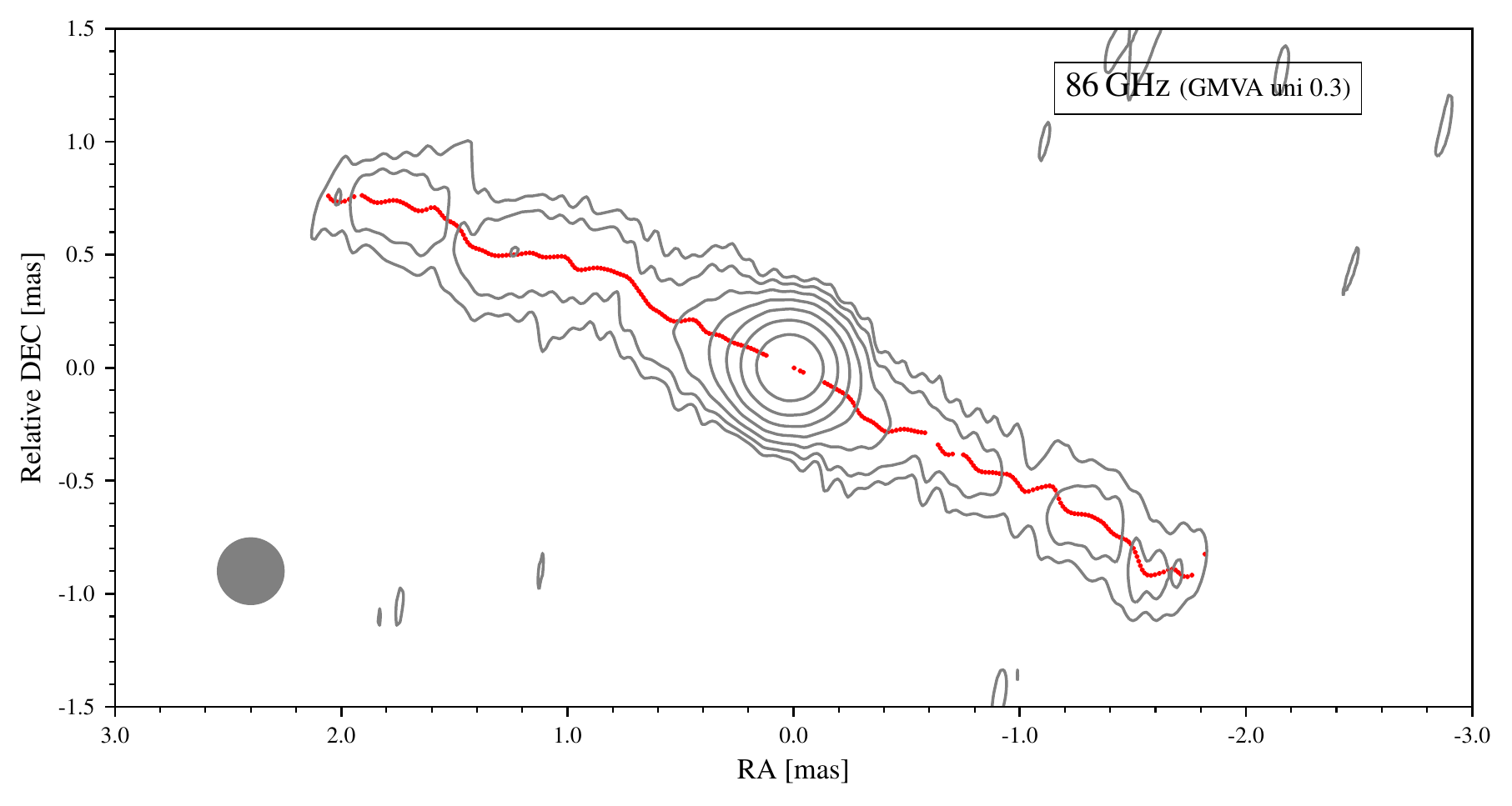}
       \caption{Uniform weighted GMVA image from 2004 restored with a circular Gaussian of $0.3\,$mas \citep{Bac16}.}
       \label{fig:gmva_2004_restored}
    \end{figure}

	 In order to ascertain the true collimation profile upstream of the break point as well as to exclude the effect of absorption and scattering, higher frequency images, which suffer less from the surrounding plasma and molecular torus, are needed. The only detection of the jets at 86\,GHz so far is from a GMVA observation from 2004 \citep{Bac16}. Given the 13 years between both observations and the close distance of this region to the central engine, those data are not included in the fitting, as it would require the assumption that no substantial changes took place over that interval. To still compare our results with the high-resolution image from 2004, we applied our ridge-line analysis to the archival data. We convolved the 86\,GHz uniform weighted image with a circular beam of $0.3\,$mas, which is slightly smaller than the major axis of the beam (see restored image in Fig.~\ref{fig:gmva_2004_restored}). 
   
   The very elliptical beam produces nonphysical, knotty structures perpendicular to the jet axis, if taking $b=\sqrt{b_\mathrm{maj}\times b_\mathrm{min}}$, as defined in Sec.~\ref{sec:ResRidge}. We assumed the 86\,GHz center to be located at the same position as the 43\,GHz center and plotted the measurements on top of the higher frequency results in Fig.~\ref{fig:RAvlba_u-w_width}. We did not include the GMVA widths into the fit. The 2004 GMVA image has a gap between the innermost, unresolved bright component and the two jets, which have significant emission at distances $z>0.2\,$mas. Hence, we only include width measurements of both jets at these larger distances. Despite the 13 years difference between both observations, the 86\,GHz widths fit remarkably well to the power-law fits. 
	 Even so the jets are only detected from 0.2\,mas distances outwards, we can still compare the collimation profile with the upper limit on the size of the unresolved central 86\,GHz core of $150\,\mu$as ($\sim 1000 R_\mathrm{S}$) \citep{Bac16}. This value is consistent with the width of the jets at the break point. Given that this is an upper limit we cannot draw further conclusions about whether the jets continue with a close-to cylindrical profile down to the jet base or transition into a parabolic expansion at closer distance. In order to estimate the initial opening angle and the collimation at distances closer than $10^3\,R_\mathrm{S}$ higher sensitivity observations with a higher North-South resolution are decisive.
	 
	\subsection{Apparent opening angle}\label{sec:apptheta}
	
	Downstream of the break point the jet is well collimated with a full apparent opening angle of $\phi_\mathrm{app}<10^\circ$. As before for the collimation profile, the western jet deviates from the overall trend at larger distances and at lower frequencies, being inconsistent with the assumption of a collimated outflow. In both jets the apparent opening angle upstream of the break increases as approaching the center. However, with increasing observing frequency the apparent opening angle decreases, being smallest at 43\,GHz. Assuming that the image at 43\,GHz is less affected by absorption and offers the highest resolution it should provide the most accurate measurement of the innermost jet opening angle and jet width. As the central region is unresolved, we exclude measurements at distances $z<0.15\,$mas, which is equivalent to half the restored, circular beam at this frequency. This leads to an apparent opening angle at a distance of $0.15\,$mas of $\phi_\mathrm{app} \sim 30^\circ$ for both jets. Taking into account a viewing angle of $i>80^\circ$ the intrinsic opening angle is given by $\phi_\mathrm{intr} = 2\,\arctan \left(\sin i\,\tan (\phi_\mathrm{app}/2)\right)$ \citep{Pus17}. Based on the large viewing angle the intrinsic angle is with $\phi_\mathrm{intr}\geq 29.6^\circ$ nearly identical with $\phi_\mathrm{app}$.
	
    \subsection{The straightness of the jets}
\begin{figure}[!hbt]
	 \centering
	 \includegraphics{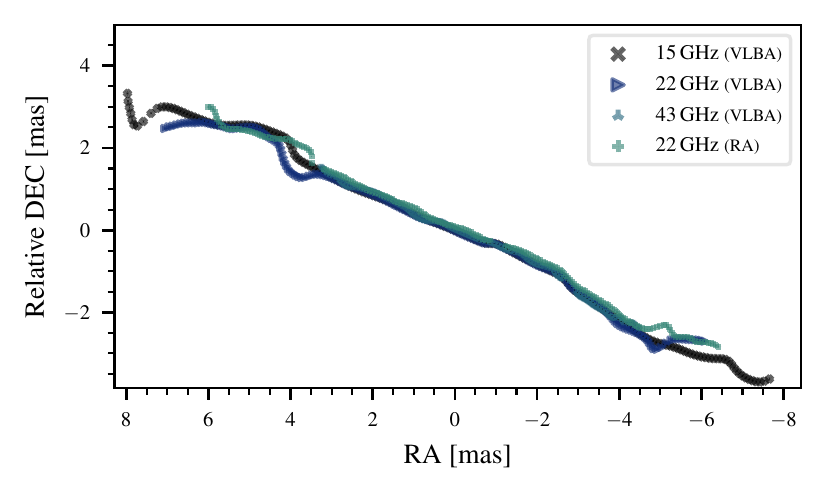}
	 \caption{Jet ridge-lines compared for Global 22\,GHz and VLBA 15, 22, 43\,GHz  observations.}\label{fig:RAvlbaridgelines}
	\end{figure}

	   Figure~\ref{fig:RAvlbaridgelines} shows all ridge-lines at frequencies of $\nu \geq 15\,$GHz plotted on top of each other. The ridge-lines are found to be exceptionally straight at all scales over the whole set of observing frequencies. This is remarkable in comparison to other AGN sources. Many radio galaxies as, for example, Cyg\,A and 3C\,84 have wide jet bases and are edge-brightened down to the closest distances to the central engine which are measurable \citep[see e.g.,][]{Gio18,Boc16b}. Bends and more complicated structures are a common feature observed for other AGN sources \citep[e.g., \object{CTA\,102}, \object{3C\,345}, \object{PKS\,0735+178}][]{Fro13,Agu06,Poe21}.  Kinematic studies of NGC\,1052 revealed higher velocities at 43\,GHz ($\beta\sim 0.5$) \citep{Bac19} than at 15\,GHz ($\beta\sim 0.2$)\citep{Lis19}. Assuming that we observe an inner, faster layer of the jets at higher frequencies this suggests jet stratification. This could be explained by assuming that the outer, slower moving layer consists of less energetic particles which are not emitting at 43\,GHz and, hence, are invisible. However, jet stratification is not visible either in the jet morphology or the spectral index distribution. This non-detection of an edge-brightened structure might very well be explained by insufficient resolution. For example, the detection of an edge-brightened structure in Centaurus~A was first discovered at the highest possible resolution obtained with the EHT \citep{Jan21}. At lower frequencies this was not observable.
	   
 There is one common feature for all images, which deviates slightly from the overall straight jet evolution. At $-2.5\,$mas the western jet changes its direction slightly towards the South to change  back to the original position angle at $-5\,$mas. There is nothing comparable in the eastern jet. This directional change coincides with the location of the break point in the collimation profile. In accordance to the previous section we assume this not to be intrinsic to the jet plasma, but to be an effect of the surrounding ionized plasma. A similar, slight jet bending in the proximity of the break is also observed at 43\,GHz images in Cyg\,A \citep{Boc16a}.
	 
	\subsection{The collimation and acceleration zone}\label{sec:colac}
	
	 The break of the broken power-law fit in both jets is located at around $10^4\,R_\mathrm{S}$ with a width at this location of $\sim 0.2\,$mas. This is about 2 orders of magnitude closer than the sphere of gravitational influence (SoI) $R_\mathrm{SoI}= 9.6\times 10^5\,R_\mathrm{S}$ for NGC\,1052 \citep{Kam20}. Hence, as the break location is clearly independent of the black hole influence, there remain two main mechanisms which can explain the observed jet properties.
	 
	 First, as suggested by numerical simulations as well as observational considerations the break most likely marks the position at which the jet transitions from being magnetically- to particle-dominated \citep{Pot15,Kom09,Kov20}. However, magnetic collimation has been claimed to not be the main collimation mechanism in extragalactic jets, as it would imply a strong toroidal magnetic field, which would make the jet very unstable \citep[][and references therein]{Miz09,Miz12,Peru17,Per19}. This is in clear contrast to the exceptional straightness of the jets.
	 
	 Secondly, the geometric transition might result from a confinement of the jet through an external pressure profile \citep{Lyu09,Bes17,Kov20}. The complex inner structure of NGC\,1052 is in favor of this explanation. NGC\,1052 is surrounded by an optically thick plasma torus which covers the inner $\sim 2\,$mas around the nucleus \citep{Saw08,Kam01} and a massive molecular torus inside a gas-poor circumnuclear disk \citep{Kam20}. The outer extent of the plasma torus is in agreement with the location of the break in the collimation profile. In addition, NGC\,1052 most likely hosts an Advection Dominated Accretion Flow (ADAF) inside an optically thick, geometrically thin accretion disk, truncated at a distance of $r_\mathrm{tr} \geq 13\,R_\mathrm{S}$ \citep{Reb18,Fal20}. This might explain the untypically close-to-cylindrical collimation profile.
	 
	 In addition, following \cite{Glo16} collimation through a disk wind could also take place, which requires relatively large wind powers. On the other hand, emission from the disk wind itself might contribute to the imaged radiation, which has the potential to prevent us from detecting the true, intrinsic jet structure. 
    
    Furthermore, the deviations of the apparent opening angle and width with frequency upstream of the break point can be explained by differential expansion, which is required by magnetic acceleration \citep{Vla04,Kom09,Kom12,Ang21}. Thus identifying the region of differential expansion with the upstream jets at distances $\lesssim10^4\,R_\mathrm{S}$ we expect the jet acceleration to take place in that region. Indeed, Fig.~6 in \cite{Bac19} shows some indication of flow acceleration within these inner $2$\,mas in both jets. This is most evident in the western jet components wj5, wj6, wj9, wj10, wj12--15 and the eastern jet components ej3-5, ej12,ej13 in Fig.~6 in \cite{Bac19}. Beyond this region, component trajectories are clearly linear, which coincides with the region in which the apparent opening angles merge for all observing frequencies within this study. The observations presented are limited by resolution and absorption processes. To further investigate a possible differential expansion accompanied by acceleration, multi-epoch observations at frequencies $>43\,$GHz have to be conducted to perform detailed kinematic studies upstream of $10^4\,R_\mathrm{S}$.
    
    In summary, external confinement and differential expansion are the most likely mechanisms for the observed jet geometry due to the deviation from a parabolic expansion and hints for jet acceleration upstream of the break point. It is especially interesting, that at 43\,GHz the profiles and opening angles are similar for eastern and western jet. In contrast to that the significant differences seen at lower frequencies makes NGC\,1052 as a double-sided jet system a very interesting target to further study the impact of the surrounding material, as the torus-accretion disk system, on the width measurement. This is not possible in one-sided jet sources. In addition, the width along both jets for the stacked 43\,GHz image suggest an even more complex structure with a second break at around $3\times 10^3\,R_\mathrm{S}$, which is more pronounced in the eastern jet. To further investigate this, higher sensitivity images at frequencies $\nu > 43\,GHz$, which resolve structures closer to the black hole, are required. With the high sensitivity of our multi-frequency and stacked images our results are the ideal starting point to study possible jet collimation mechanism and the impact of a change in the outer pressure profile in greater detail by comparing with and guiding high-resolution 3D numerical simulations.
    	 
    \section{Summary}
    \label{sec:summary}
    
    We have presented a detailed study of the ridge-line and collimation profile of NGC\,1052 over a frequency range from $1.5$\,GHz to $86$\,GHz observed with the VLBA and a large global array supporting a RadioAstron observation. Further, we compare the low radio power AGN NGC\,1052 with powerful FR\,I and FR\,II sources (e.g., 3C\,84, M\,87, and Cygnus~A). The comparison between low- and high-power jets is extremely important to test the universality of the jet phenomenon. Our main conclusions are summarized as follows:
    \begin{compactitem}
        \item In accordance with previous observations we find a remarkably straight, low velocity double-sided jet. There is no hint for any double ridge-line depicting an edge brightened structure as observed in other nearby AGN with highly relativistic jets as, e.g., \object{Cyg\,A}, \object{M\,87}, or \object{3C\,84}. 
        \item There is a change from a nearly cylindrical to conical outflow at a break point of $\sim 2\,$mas or $10^4\,R_\mathrm{S}$, slightly different for eastern and western jet. This is much closer than the SoI of the black hole and we do not find any trace of a parabolic jet shape. NGC\,1052 most likely hosts an ADAF inside a truncated, thin disk which is surrounded by a dense torus \citep{Kam01,Reb18,Kam20}. This makes a scenario in which the jets are confined by an outer pressure gradient the most likely one.
        \item At lower frequencies $<15\,$GHz there are significant differences between the eastern, approaching jet and the western, receding jet. On the one hand, the eastern jet is well described by displaying a conical expansion outside the break point over all frequencies. On the other hand, width measurements in the western jet at 1.5\,GHz and 5\,GHz deviate strongly from the conical expansion showing a much wider jet as is expected, suggesting scattering from the dense molecular torus found by 
        ALMA observations \citep{Kam20}.
        \item Width measurements of archival $86$\,GHz data from 2004 are in agreement with the cylindrical jet shape upstream of the break point. However, the limited sensitivity and unresolved central component prevent us from drawing conclusions at distances $<10^3\,R_\mathrm{S}$ to the central engine.
        \item With increasing frequency the apparent opening angle decreases. In combination with a flow acceleration within the inner 2\,mas, which is indicated in Fig.\,6 in \cite{Bac19}, this can be explained by differential expansion, which is required by magnetic acceleration. We do not detect an edge-brightened structure. This can be explained by insufficient resolution as it is the case for Centaurus~A, which only revealed edge-brightening at high frequency observations with the EHT.
        \item At frequencies $>15\,$GHz the single-epoch observations reveal significant differences between eastern and western jet, which is not observed by \cite{Nak20}. However, by fitting stacked images the upstream profile is with $k_\mathrm{u}=0.19$ close-to cylindrical for both jets and the downstream profiles are with $k_\mathrm{u}=0.83$ and $k_\mathrm{u}=0.98$ close to conical for eastern and western jet, respectively. This suggests that the full jet width can only be probed by stacked images. Still, the $15$\,GHz and $22$\,GHz stacked images reveal larger widths in the western jet in comparison to the eastern jet. Moreover at distances $<0.2\,$mas the 43\,GHz widths get smaller and their profile of $k_\mathrm{u}=0.30\pm0.03$ and $k_\mathrm{u}=0.22\pm0.02$ for the eastern and the western jet, respectively, deviate from the cylindrical profile observed at $15$ and $22$\,GHz.
    \end{compactitem}
    In order to investigate the jet shape upstream of the break point in more detail high frequency, multi-epoch observations at frequencies $\geq 86\,$GHz with a higher North-South resolution are decisive. The same holds for estimation of the initial opening angle.
    
     \begin{acknowledgements} 
     The authors would like to thank the anonymous referee for his/her useful comments. We thank Michael Janssen for his valuable comments to the manuscript. We thank Gabriele Bruni and Tuomas Savolainen for their strong support in the correlation of the RadioAstron observation.
     The RadioAstron project is led by the  Astro Space Center of the Lebedev Physical Institute  of the Russian Academy of Sciences and the Lavochkin Scientific and Production Association under a contract with the State Space Corporation ROSCOSMOS, in collaboration with partner organizations in Russia and other countries. Partly based on observations performed with radio telescopes of IAA RAS (Federal State Budget  Scientific Organization Institute of Applied  Astronomy of Russian Academy of Sciences). The European VLBI Network is a joint facility of independent European, African, Asian, and North American radio astronomy institutes. Partly based on observations with the 100-m telescope of the MPIfR (Max-Planck-Institut für Radioastronomie) at Effelsberg. Results of optical positioning measurements of the Spektr-R spacecraft by the global MASTER RoboticNet \citep{Lip10}, ISON collaboration, and Kourovka observatory were used for spacecraft orbit determination in addition to mission facilities. The National Radio Astronomy Observatory is a facility of the National Science Foundation operated under cooperative agreement by Associated Universities, Inc. The data were correlated at the DiFX correlator \citep{Del11,Bru16} of the MPIfR at Bonn.  This research has made use of NASA’s Astrophysics Data System. This research has made use of the NASA/IPAC Extragalactic Database (NED), which is operated by the Jet Propulsion Laboratory, California Institute of Technology, under con-tract with the National Aeronautics and Space Administration. This research has made use of Astropy, a community-developed core Python package for Astronomy \citep{astropy13,astropy18}, the scikit-image bibliography \citep{scikit}, scipy \citep{scipy}, and the ODR package, based on \citep{Bog90}. MP acknowledges support by the Spanish Ministry of Science through Grants PID2019-105510GB-C31, PID2019-107427GB-C33, and from the Generalitat Valenciana through grant PROMETEU/2019/071. CMF is supported by the black hole initiative at Harvard University, which is supported by a grant from the John Tempelton Fundation.
    \end{acknowledgements}
    
 \bibliographystyle{aa}
 \bibliography{bibliography}
   \end{document}